\documentclass{cpcauth}

\usepackage{amssymb}

\begin{document}

\begin{frontmatter}



\title{Density of Partition Function Zeroes and Phase Transition Strength}


\author[label1]{Wolfhard Janke}
and 
\author[label2]{Ralph Kenna}

\address[label1]{Institut f\"ur Theoretische Physik,
               Universit\"at Leipzig, 
               Augustusplatz 10/11, 
               04109 Leipzig, 
               Germany}

\address[label2]{School of Mathematics, 
                 Trinity College Dublin, Ireland \\
and School of Mathematical and Information Sciences,
Coventry University, CV1 5FB, England}

\begin{abstract}
A new method to extract the density of partition function zeroes
(a continuous function) from their distribution for finite lattices 
(a discrete data set) is presented. This allows direct determination 
of the order and strength of phase transitions numerically. Furthermore,
it enables efficient distinguishing between first and second order 
transitions, elucidates crossover between them and illuminates the 
origins of finite-size scaling.
The  efficacy of the technique is demonstrated by its application to
a number of models in the case of Fisher zeroes and to the $XY$
model in the case of Lee-Yang zeroes.
\end{abstract}

\end{frontmatter}

\section{Introduction}
\label{Introduction}

Phase transitions are phenomena common to a wide range of disciplines,
from physics to biology, economics and even sociology.
Examples include neural networks, protein folding, melting, magnetism,
stock market crashes 
and the deconfinement transition in the early universe.

In statistical physics, in particular, one is interested in the 
determination of the location, order and strength of phase transitions.
First order transitions  involve the
coexistence of two or more distinct phases and 
are characterised by a discontinuity in the first
derivative of the free energy
corresponding to the finite amount of energy needed to transform
from one phase to the other.
For temperature driven transitions, this
discontinuity is the latent heat $\Delta e$ 
and is a measure of the transition strength.
For second order transitions there is no such coexistence. 
Instead, thermodynamic quantities such as the correlation length 
and specific heat diverge.
Such divergences are characterised by critical exponents $\nu$ and
$\alpha$, which is a measure of the strength of the transition.

The computational approach to the study of phase transitions consists of two
steps -- the gathering of data in the form of a Monte Carlo
simulation followed by a numerical analysis of appropriate quantities.
The first of these is restricted to systems of finite size. 
Phase transitions, however, require an infinite number of available states 
for their occurrence. The second step in the numerical approach
is an extrapolation to infinite volume.
Traditional techniques involve the 
finite-size scaling (FSS) study of thermodynamic functions.
An increasingly popular alternative approach is, however,
 the use of zeroes of the partition 
function.

Let $t=T/T_c-1$ be the reduced temperature and $h$ the external field.
For a $d$-dimensional system of linear extent $L$, 
the FSS of the $j^{\rm{th}}$ complex partition function zero
(for large $j$) is given by \cite{IPZ}
\begin{equation}
 t_j(L) \sim \left( {j}/{L^d} \right)^{1/\nu d}
\quad, 
\label{FSSa}
\end{equation}
or
\begin{equation}
 h_j(L) \sim \left({j}/{L^d}\right)^{(d+2-\eta)/2d}
\quad .
\label{FSSb}
\end{equation}
Here $\eta$ is the anomalous dimension and $h=0$ in the first
formula (where the zeroes are called Fisher zeros)
while $t=0$ in the second (where the zeroes are Lee-Yang zeroes).
The standard approach to FSS of zeroes is to fix the index to
$j=1$ and extract an estimate for the critical
exponents from a range of lattice sizes.
It has, however, long been known that using more than one index
could provide more information. However, since (\ref{FSSa}) and
(\ref{FSSb}) are inexact,
this has been prohibitive. In particular, the  extraction of
the density of zeroes (a continuous function) from their (discrete)
distribution for a finite and numerically accessible lattice 
has been considered prohibitively difficult or even impossible
\cite{Martin}.
In recent years, however, there have been some attempts to 
overcome the difficulties involved
 \cite{density}.
In view of the increasing importance attached to this approach, we
suggest an appropriate way this should be done \cite{us}.

%%%%%%%%%%%%%%%%%%%%%%%%%%%%%%%%%%%%%%%%%%%%%%%%%%%%%%%%%%%%%%%%%
% main text
\section{Density of Zeroes}
\label{theory}

The partition function for finite $L$ is
$ Z_L(z) \propto  \prod_{j}{\left(z-z_j(L)\right)}$,
where $z$ is an appropriate function of temperature
or field.
%%%%(Here the focus is on zeroes in the complex temperature plane,
%%%%although the methods extend to the  external field plane too.)
We assume the zeroes, $z_j$, are 
on a line impacting on to the real axis 
at the critical point, $z_c$. 
Parameterising zeroes on this  line by
$z_j=z_c+r_j \exp{(i \varphi)}$ we may define 
the density of zeroes as
$
 g_L(r) = L^{-d} \sum_{j} \delta(r - r_j(L))
$.
The cumulative distribution function of zeroes is then
$
 G_L(r)
 =
 \int_0^r{ g_L(s) d s}
$
which is $j/L^d$ if $ r \in (r_j,r_{j+1})$.
At a zero we assume the cumulative density 
is given by the average  \cite{LY,2j-1}
\begin{equation}
 G_L(r_j) =  (2j-1)/2L^d
\quad .
\label{III}
\end{equation}

In the thermodynamic limit and for a phase transition of 
first order this integrated density of zeroes is, in fact \cite{LY},
\begin{equation}
 G_\infty(r) = g_\infty(0) r 
\quad ,
\label{1st}
\end{equation}  
so that the density is non-vanishing at the real axis.
The slope at the origin in (\ref{1st})  is related to the latent heat
in the Fisher case
(or magnetization in the Lee-Yang case) 
 via \cite{LY}
\begin{equation}
 g_\infty(0) \propto  \Delta e 
\quad .
\label{1stl}
\end{equation}

For a second order transition
the corresponding expressions for Fisher and Lee-Yang zeroes are
 \cite{Abe}
\begin{equation}
 G_\infty(r) \propto  r^{2-\alpha} 
\quad {\rm{or}} \quad 
G_\infty(r) \propto  r^{2d/(d+2-\eta)}
\quad ,
\label{2nd}
\end{equation}
respectively.

Thus while the scaling behaviour of the position of the first few
zeroes
in the complex temperature plane can  be used to identify  
$\nu$ via (\ref{FSSa}), the density of zeroes gives the strength of
the transition. A plot of $G_L(r_j)$ against $r_j(L)$
should ({\em{i}\/}) go through the origin, ({\em{ii}\/}) display $L$-- and
$j$-- collapse and ({\em{iii}\/}) reveal the order and strength of the 
phase transition by its slope near the origin.

In  (\ref{III}), $r_j$ may be taken to be the imaginary part of the 
position of the $j^{\rm{th}}$ zero. Equating (\ref{III}) to 
(\ref{2nd}) in the second order Fisher case, gives $r_j(L) \sim
L^{-1/\nu}$. This is the usual FSS formula for fixed index
Fisher zeroes. Similarly, in the Lee-Yang case, one recovers
the fixed index FSS formula $h_j(L) \sim L^{-(d+2-\eta)/2}$.
Also, equating (\ref{III}) to (\ref{1st}) 
gives $r_j(L) \sim L^{-d}$, explaining the usual
identification of $\nu$ with 
$1/d$ for a first order temperature driven phase transition. 
Therefore, traditional FSS emerges quite 
naturally from this density approach.

%%%%%%%%%%%%%%%%%%%%%%%%%%%%%%%%%%%%%%%%%%%%%%%%%%%%%%%%%%%%%%%%%
% main text
\section{Applications}
\label{applications}

To demonstrate  our approach, 
we perform fits to the cumulative density
of zeroes for a number of different models in statistical physics
and in lattice field theory \cite{us}.
Here we present the results for the Fisher zeroes
for two models, one from each field. Furthermore, we discuss the
Lee-Yang zeroes of the two-dimensional $XY$ model.

Allowing for first or second order behaviour, the cumulative density
should behave as 
\begin{equation}
 G(r) = a_1 r^{a_2} + a_3
\quad ,
\label{gen}  
\end{equation}
where we also allow for an additional parameter $a_3$
which should be zero for a good fit.
In fact, a non-zero value of $a_3$ indicates the absence of a 
phase transition, for, $a_3 > 0$ means the zeroes have already crossed
the real axis (the situation in the broken phase) while
$a_3<0$ means the zeroes have not yet reached the real axis 
(the symmetric phase).
For Fisher zeroes, a first order phase transition is 
indicated if $a_2 \sim 1$ for small $r$, in which case
the latent heat is proportional to the slope $a_1$.
A value of $a_2$ larger than
$1$ signals a second order 
transition whose  strength is 
given by $\alpha = 2-a_2$.
Note that $\alpha$ can be measured {\em{directly}} using
this method while traditional FSS only allows the measurement of the 
ratio $\alpha / \nu$.

%%%%%%%%%%%%%%%%%%%%%%%%%%%%%%%%%%%%%%%%%%%%%%%%%%%%%%%%%%%%%%%%%%%%%%%%
~\\
\noindent
{\em{The $d=2, q=10$ Potts Model:}}
%%%%%%%%%%%%%%%%%%%%%%%%%%%%%%%%%%%%%%%%%%%%%%%%%%%%%%%%%%%%%%%%%%%%%%%%
The first six Fisher zeroes for the 
two-dimensional 
$10$--state Potts model for lattice sizes 
$L=4$--$64$ are listed in \cite{Vi91}.
A traditional FSS analysis applied to the first zero
for large lattices
provides  evidence for $\nu = 1/d$ and 
hence a first order phase transition. However, the determination of 
the lattice size
above which FSS sets in is,
by necessity,  somewhat arbitrary.
Indeed, when one extends the analysis to higher index zeroes 
one finds that when corrections are 
ignored,
no two-parameter fit gives an acceptable result.

\begin{figure}[t]
\vspace{5cm}
%\special{psfile=d=2.q=10.ps angle=-90  hoffset=-10 voffset=170 
%                                           hscale=30 vscale=30}
\includegraphics{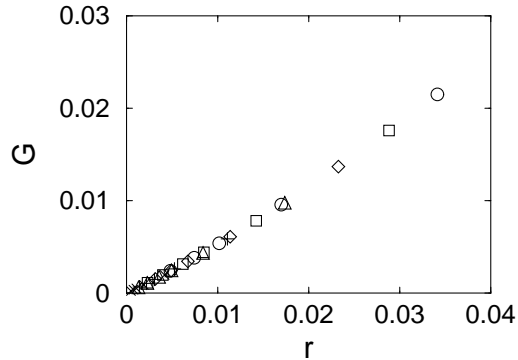}
\caption[a]{Distribution of zeroes for the 
$d=2$, $q=10$ Potts model (for $L=16$--$64$) which has a 
first order phase transition. The symbols 
$\times$,
$+,\bigtriangleup,\diamond,\put(1,0){\framebox(5,5)}~~$, and 
$\put(4,3){\circle{5}}~~,$
correspond to $j=1,2,3,4,5$,  
and $6$, respectively.}
\label{fig:d=2.q=10}
\end{figure}
Our analysis of the density of zeroes begins with
Fig.~\ref{fig:d=2.q=10}. The
excellent data collapse for various $L$ and $j$
indicates that (\ref{III})
is the correct form for the density of zeroes.
Fitting (\ref{gen}) to the $L=16$--$64$, $j=1$--$4$ 
data points gives 
$a_2=1.10(1)$ and $a_3= 0.00004(1)$,
indicating a first 
order phase transition.
Fixing $a_3=0, a_2=1$
and applying a single-parameter fit close to the origin 
yields a slope corresponding to latent heat
$\Delta e = 0.698(2)$ 
which compares well with the exact value $0.6961$.

%%%%%%%%%%%%%%%%%%%%%%%%%%%%%%%%%%%%%%%%%%%%%%%%%%%%%%%%%%%%%%%%%%%%%%%%
~\\
\noindent
{\em{The $d=4$, Abelian Surface Gauge Model:}}
%%%%%%%%%%%%%%%%%%%%%%%%%%%%%%%%%%%%%%%%%%%%%%%%%%%%%%%%%%%%%%%%%%%%%%%%
This is a model dual to the $d=4$ Ising
model, which, up to logarithmic corrections has mean field
critical exponents \cite{KeLa93}. One therefore expects the
surface gauge model also to be characterised by
mean field exponents with $\alpha=0$ and $\nu = 1/2$.

The first two Fisher 
zeroes for lattices of size $L=3$ to $12$ are listed in
 \cite{BaVi94} where a 
conventional analysis applied to the first index zero
yields the best estimate of $\nu = 0.469(17)$ 
from the two largest lattices. Inclusion of the smaller
lattices worsens the fit driving $\nu$ away from $1/2$.
Also, a bimodal structure in the energy histograms 
appears as a spurious indication 
of a first order transition \cite{BaVi94}.

\begin{figure}[t]
\vspace{5cm}
%\special{psfile=d=4.surface.gauge.ps angle=-90  hoffset=-14 voffset=170 
%                                           hscale=31 vscale=28}
\includegraphics{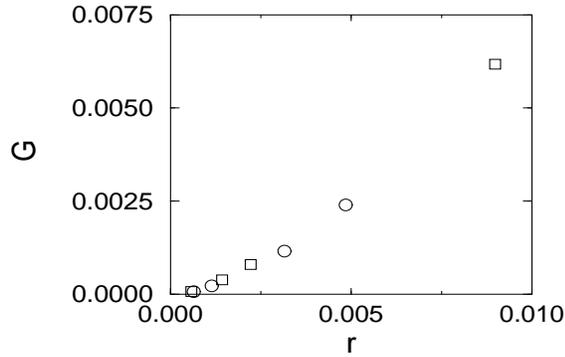}
\caption[a]{Distribution of zeroes for the four-dimensional Abelian
surface gauge model which has a  second order transition.
The symbols $\put(1,0){\framebox(5,5)}~~$ and $\put(4,3){\circle{5}}~~~$
correspond to the $j=1$ and $j=2$ index zeroes, respectively.}
\label{fig:d=4.surface.gauge}
\end{figure}

A fit of the data to (\ref{gen}) yields $a_2$ 
incompatible with unity
(see Fig.~\ref{fig:d=4.surface.gauge}), with
a fit near the origin yielding $a_2=1.90(9)$.
This corresponds to $\alpha = 0.10(9)$, compatible with zero.
Note  that only the region near the origin in 
Fig.~\ref{fig:d=4.surface.gauge} is of interest.
The slope there is not
compatible with a first order transition and the bimodal structure
of the energy histograms 
observed in \cite{BaVi94} can only be an unexplained
finite-size effect.

\begin{figure}[t]
\vspace{5cm}
%\special{psfile=d=2.xy.ps angle=-90  hoffset=-14 voffset=170 
%                                           hscale=31 vscale=30}
\includegraphics{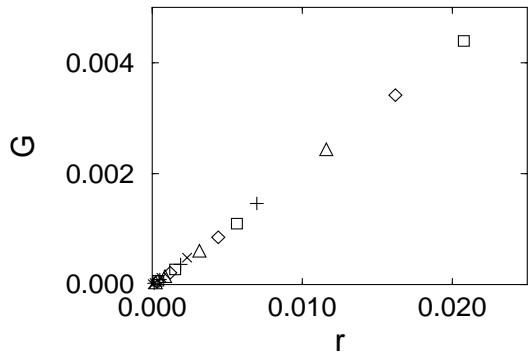}
\caption[a]{Distribution of Lee-Yang zeroes for the two-dimensional $XY$
model at $\beta = 1.113$ (for $L=32-256$).
The symbols 
$\times$,
$+,\bigtriangleup,\diamond$, and $\put(1,0){\framebox(5,5)}~~~$
correspond to $j=1,2,3,4$,  
and $5$, respectively.}
\label{fig:d=2.xy}
\end{figure}

%%%%%%%%%%%%%%%%%%%%%%%%%%%%%%%%%%%%%%%%%%%%%%%%%%%%%%%%%%%%%%%%%%%%%%%%
~\\
\noindent
{\em{The $d=2$  $XY$ Model:}}
%%%%%%%%%%%%%%%%%%%%%%%%%%%%%%%%%%%%%%%%%%%%%%%%%%%%%%%%%%%%%%%%%%%%%%%%
Here we demonstrate that the density technique is also applicable
in the Lee-Yang case.
The first few Lee-Yang zeroes for the $d=2$ $XY$ model 
with $L=32$--$256$ were determined for
various temperatures in Ref.~\cite{KeIr97}. 
Figure~\ref{fig:d=2.xy} depicts the distribution of 
these zeroes for $\beta = 1/k_B T = 1.113$ which is the critical point
(here $k_B$ is the Boltzmann constant). From (\ref{2nd}), and with 
$\eta = 1/4$ (the value of the anomalous dimension in this model),
one expects $G_L(r_j) \sim r_j^{16/15}$, where $r_j=h_j$. A three-parameter
fit to (\ref{gen}) gives $a_3 = 0$ indicating that criticality has 
indeed been reached at this temperature. A two-parameter fit now yields 
$a_2=1.06(1)$, compatible with expectation. 
One notes that logarithmic corrections are present in this model
as shown in \cite{KeIr97,Ja97}.

%%%%%%%%%%%%%%%%%%%%%%%%%%%%%%%%%%%%%%%%%%%%%%%%%%%%%%%%%%%%%%%%%%%%%%%%
%%%%%%%%%%%%%%%%%%%%%%%%%%%%%%%%%%%%%%%%%%%%%%%%%%%%%%%%%%%%%%%%%%%%%%%%
\section{Conclusions}
%%%%%%%%%%%%%%%%%%%%%%%%%%%%%%%%%%%%%%%%%%%%%%%%%%%%%%%%%%%%%%%%%%%%%%%%
%%%%%%%%%%%%%%%%%%%%%%%%%%%%%%%%%%%%%%%%%%%%%%%%%%%%%%%%%%%%%%%%%%%%%%%%
\label{conclusions}  

We have presented a new method to extract the (continuous)
density of zeroes from
(discrete) 
finite-size data and demonstrated how this can be used to distinguish
between phase transitions of first and second order as well as
 to measure their
strengths. The method meets with a high degree of success in statistical
physics and lattice field theory 
and lends new insights into the
origins of finite-size scaling.

\end{document}